\documentclass[11pt,a4paper]{article}
\usepackage[utf8]{inputenc}
\usepackage[T1]{fontenc}
\usepackage{lmodern}
\usepackage{microtype}
\usepackage[margin=2.5cm]{geometry}
\usepackage{setspace}
\usepackage{parskip}
\usepackage{graphicx}
\usepackage{booktabs}
\usepackage{enumitem}
\usepackage{hyperref}
\hypersetup{colorlinks=true,linkcolor=blue,citecolor=blue,urlcolor=blue}
\usepackage{titling}
\usepackage{fancyhdr}
\usepackage{caption}
\usepackage{amsmath,amssymb}
\usepackage{natbib}
\usepackage{enumitem}
\usepackage{etoolbox}
\AtBeginEnvironment{thebibliography}{%
  \setlength{\itemsep}{0.5pt}   
  \setlength{\parskip}{0pt}     
  \setlength{\parsep}{0pt}      
  \setlength{\bibsep}{0pt}      
}

\pretitle{\vspace*{-1cm}\begin{center}\LARGE\bfseries}
\posttitle{\par\end{center}\vskip 0.5em}
\preauthor{\begin{center}\large}
\postauthor{\end{center}}
\predate{\begin{center}\large}
\postdate{\end{center}}

\usepackage[dvipsnames]{xcolor}

\begin{document}

\begin{titlepage}
  \centering
  \vspace*{0.1cm}
  {\Huge\bfseries Expanding Horizons \\[6pt] \Large Transforming Astronomy in the 2040s \par}
  \vspace{0.5cm}

  {\LARGE \textbf{Time-Domain Multi-Messenger Astronomy in the 2040s: EM Follow-up of LGWA Sources}\par}
  \vspace{0.5cm}

  \begin{tabular}{p{4.5cm}p{10cm}}
    \textbf{Scientific Categories:} & Multi-messenger astrophysics, Time-domain  astrophysics, Gravitational wave sources, binary evolution\\
    \\
    \textbf{Submitting Author:} & Ferdinando Patat \\
    & European Southern Observatory \\
    & fpatat@eso.org\\
    \\
    \textbf{Contributing authors:}
& Silvia Piranomonte, INAF-Observatory of Rome, \\
& Stefano Benetti, INAF-Observatory of Padova, \\ 
& Alessandro Bonforte, INGV-Catania,\\
& Roberto Della Ceca, INAF-Observatory of Brera,\\
& Gianluca Di Rico, INAF-Observatory of Abruzzo,\\
& Alessandro Frigeri, INAF-Bologna, \\
& Jan Harms, Gran Sasso Science Institute, \\
& Marco Olivieri, INGV-Bologna, \\
& Andrea Perali, University of Camerino, \\
& Paola Severgnini, INAF-Observatory of Brera and\\
& Angela Stallone, INGV-Bologna, \\
& on behalf of the LGWA Collaboration\\
  \end{tabular}

  \vspace{1cm}

  \textbf{Abstract}
      
  \vspace{0.5em}
  \begin{minipage}{0.9\textwidth}
    \small
    The coming decades will see gravitational-wave (GW) astronomy expand decisively into the mHz–Hz frequency range, opening access to a population of compact binaries that are currently invisible or only detectable moments before merger. The Lunar Gravitational Wave Antenna (LGWA) concept is designed to probe this gap, enabling continuous observation of compact binaries over months to years prior to coalescence, and detecting sources inaccessible to both space-based mHz detectors and current ground-based $>$10 Hz facilities. This new GW window fundamentally alters the landscape of time-domain multi-messenger astronomy. Rather than reacting to mergers after the fact, LGWA enables predictive, scheduled electromagnetic (EM) follow-up, transforming how compact-object mergers, their environments, and their astrophysical channels are studied.
    However, fully exploiting LGWA discoveries requires EM capabilities that do not exist today and are unlikely to be available by the 2030s, particularly for wide-area, rapid, spectroscopically rich follow-up at optical and near-infrared wavelengths. This White Paper identifies the key science cases enabled by LGWA that motivate new ground-based capabilities in the 2040s.
  \end{minipage}

\end{titlepage}


\section{Introduction and Background}
\label{sec:intro}

The Lunar Gravitational Wave Antenna opens a new observational window on the GW sky by targeting the poorly explored frequency band between the mHz regime of space-based detectors and the tens-of-hertz band of terrestrial interferometers. LGWA will be sensitive to GWs in the $\sim$1 mHz–1 Hz range, bridging the gap between missions such as LISA and current and next-generation ground-based observatories.\cite{Harms2021} This unique frequency coverage enables the long-term tracking of compact binaries well before merger, providing early-warning timescales of days to months for neutron star and black hole systems. As a result, LGWA fundamentally transforms multi-messenger astronomy by enabling predictive, scheduled EM follow-up and long-baseline time-domain studies that are not possible with existing or currently planned facilities.\cite{LGWAWP} 

\section{Open Science Questions in the 2040s}

By opening access to the largely unexplored decihertz gravitational-wave band, LGWA will enable a broad range of science investigations \cite{DeciHzScience}. A comprehensive discussion of the LGWA science case is presented in the LGWA White Paper \cite{LGWAWP}. The science cases highlighted below emphasize regimes where predictive, long-baseline multi-messenger observations are essential, and where current or planned EM facilities are insufficient. 

\subsection{Pre-Merger EM Evolution of Neutron Star Binaries}

LGWA will detect binary neutron star (BNS) and neutron star–black hole (NSBH) systems long before merger, potentially providing days to months of early warning. This enables a fundamentally new observational regime:

\begin{itemize}[parsep=2pt, topsep=2pt]
\item Construction of deep, contemporaneous reference images immediately before merger, minimizing host subtraction systematics.
\item Minute-to-hour cadence EM observations starting at merger time, capturing the earliest phases of kilonova emission.
\item Time-resolved color and spectral evolution of the fastest, hottest ejecta components, which are poorly constrained today due to delayed discovery.
\end{itemize}

{\bf Limitation today:}
While wide-field imagers and large telescopes exist, no current or planned facility combines wide-field coverage of large GW localization regions, rapid scheduling flexibility, and immediate multiplexed spectroscopy of large numbers of candidate counterparts. As a result, early kilonova phases are typically missed or sparsely sampled, and spectroscopic confirmation is delayed or incomplete.

\subsection{Host-Galaxy Identification and Environmental Diagnostics Before Merger}

Because LGWA tracks sources over extended periods, it enables progressive localization refinement and host-galaxy ranking before any explosive EM signal appears. This makes it possible to study:
\begin{itemize}[parsep=2pt, topsep=2pt]
\item The galactic environments of compact-object mergers (e.g. star-forming regions, globular clusters, galactic nuclei).
\item The role of AGN disks or dense nuclear environments in facilitating mergers.
\item Pre-merger variability or subtle EM perturbations in candidate host systems.
\end{itemize}

{\bf Limitation today:}
Such studies require repeated wide-field spectroscopic monitoring of large galaxy samples within evolving localization volumes. Existing spectroscopic facilities lack the combination of field of view, multiplexing, cadence, and ToO agility needed for this systematic, pre-merger approach.

\subsection{Intermediate-Mass and Exotic Compact Binary Populations}

LGWA will be sensitive to sources such as:

\begin{itemize}[parsep=2pt, topsep=2pt]
\item Intermediate-mass black hole binaries,
\item Eccentric and dynamically formed systems,
\item Hierarchical mergers and triples.
\end{itemize}

These sources probe poorly understood formation channels and environments. Even if intrinsically EM-faint, their interaction with surrounding matter (gas, stars, or nuclear regions) may produce subtle, time-dependent EM signatures.

{\bf Limitation today:}
Detecting and characterizing these signals requires deep, high-cadence monitoring of dense stellar and nuclear regions, systematic spectroscopic searches for transient or slowly evolving features, long-term, coordinated follow-up campaigns over months to years. Such programmes are currently infeasible at scale due to limited spectroscopic survey speed and coordination constraints.

\subsection{Strongly Lensed and Multi-Band GW Events}

In the 2040s, LGWA detections will be combined with other GW observatories, enabling identification of strongly lensed GW events and multi-band detections of the same source. EM follow-up will be critical to:

\begin{itemize}[parsep=2pt, topsep=2pt]
\item Identify and model lensing galaxies,
\item Confirm multiple GW “images” via host association,
\item Constrain cosmological and astrophysical parameters.
\end{itemize}

\textbf{Limitation today:}
Rapid, wide-field, high-resolution imaging and spectroscopy of multiple candidate hosts is not achievable within the necessary time windows using existing facilities.

\subsection{Type Ia SNe: Progenitors, Explosion Channels and Early-Time Physics}

Type Ia supernovae (SNe Ia) are cornerstone tools for precision cosmology, yet their progenitor systems and explosion mechanisms remain poorly understood. Competing scenarios—including single-degenerate systems, double-degenerate WD mergers, and dynamically driven detonations—predict distinct pre-explosion binary configurations and environments, but current EM observations alone are insufficient to unambiguously discriminate among them.

LGWA will directly probe compact white dwarf (WD) binaries in the mHz–Hz GW band, providing access to the immediate progenitor population of SNe Ia. In particular, LGWA will be sensitive to:

\begin{itemize}[parsep=2pt, topsep=2pt]
\item Double WD systems approaching merger,
\item Eccentric or dynamically formed binaries,
\item Long-lived inspirals that may lead to delayed detonations.
\end{itemize}

Crucially, LGWA detections will enable predictive identification of potential SN Ia progenitors, in some cases years to months before explosion. This transforms the study of SNe Ia from a purely reactive enterprise into a predictive, multi-messenger investigation.

EM follow-up of LGWA-identified progenitors enables science that is inaccessible today: a) pre-explosion monitoring of candidate systems to search for mass transfer, accretion signatures, or circumstellar material; b) immediate, minute-to-hour cadence observations at explosion time, capturing shock interaction, companion signatures, or surface detonations that fade rapidly; c) early-time spectroscopy to constrain composition, asymmetry, and ignition physics before radiative diffusion erases key information. Beyond explosion physics, the combination of LGWA and EM observations will allow calibration of SN Ia populations based on physical progenitor properties rather than empirical light-curve relations, with far-reaching implications for cosmology and stellar evolution.

\textbf{Limitation today:} Current facilities lack both the forewarning and the coordinated time-domain infrastructure required to execute such programs systematically. Even in the era of large optical surveys, SNe Ia are discovered hours to days after explosion, when the most diagnostic signals are already lost. By contrast, LGWA enables targeted, high-cadence observations of known progenitor systems at the moment of explosion, providing a decisive path toward resolving the long-standing question of SN Ia progenitors.

\subsection{Other science cases}

Facilities capable of addressing LGWA follow-up needs would also transform core-collapse supernova physics, tidal disruption events, AGN variability and nuclear transients, and compact-object demographics.

\section{Technology and Data Handling Requirements}
\label{sec:tech}

Fully exploiting this scientific potential places stringent requirements on ground-based EM capabilities. In particular, rapid and highly multiplexed spectroscopic follow-up is required to identify and physically characterise counterparts over large sky areas, with spectroscopy (rather than imaging alone) constituting the primary bottleneck. In parallel, high-cadence, wide-field optical and NIR coverage is needed on timescales relevant for early counterpart evolution. Finally, the science cases outlined in the previous sections imply data handling and operational frameworks capable of coordinating and classifying hundreds to thousands of transient candidates per event, with seamless integration of multi-messenger triggers.


\bibliography{references}  

@article{Harms2021,
doi = {10.3847/1538-4357/abe5a7},
url = {https://doi.org/10.3847/1538-4357/abe5a7},
year = {2021},
month = {mar},
publisher = {The American Astronomical Society},
volume = {910},
number = {1},
pages = {1},
author = {Harms, Jan and others},
title = {Lunar Gravitational-wave Antenna},
journal = {The Astrophysical Journal}
}

@article{LGWAWP,
doi = {10.1088/1475-7516/2025/01/108},
url = {https://doi.org/10.1088/1475-7516/2025/01/108},
year = {2025},
month = {jan},
publisher = {IOP Publishing},
volume = {2025},
number = {01},
pages = {108},
author = {Ajith, Parameswaran and others},
title = {The Lunar Gravitational-wave Antenna: mission studies and science case},
journal = {Journal of Cosmology and Astroparticle Physics}
}

@article{DeciHzScience,
doi = {10.1088/1361-6382/abb5c1},
url = {https://doi.org/10.1088/1361-6382/abb5c1},
year = {2020},
month = {oct},
publisher = {IOP Publishing},
volume = {37},
number = {21},
pages = {215011},
author = {Sedda, Manuel Arca and others},
title = {The missing link in gravitational-wave astronomy: discoveries waiting in the decihertz range},
journal = {Classical and Quantum Gravity}
}

\end{document}